\documentclass{elsart3p}

\usepackage{graphics,graphpap,amssymb,psfrag,subfigure,picinpar}
\usepackage{graphicx}

\begin{document}

\begin{frontmatter}

\title{Sharp Superconductor-Insulator Transition in Short Wires}
 \author[label1]{Dganit Meidan},
 \author[label1]{Yuval Oreg},
 \author[label2]{Gil Refael}, and
  \author[label3]{Robert A. Smith}

  \address[label1]{Department of Condensed Matter Physics, Weizmann Institute of Science, Rehovot, 76100, Israel}
 \address[label2]{Department of Physics, California Institute of Technology,
Pasadena, California 91125, USA}
 \address[label3]{School of Physics and Astronomy, University of Birmingham, Edgbaston, Birmingham B15 2TT, England}

\begin{abstract}
Recent experiments on short MoGe nanowires show a sharp
superconductor-insulator transition tuned by the normal
state resistance of the wire, with a critical resistance of $R_c\approx R_Q=
h/(4e^2)$. These results are at odds with a broad range of
theoretical work on Josephson-like systems that predicts a smooth
transition, tuned by the value of the resistance that shunts the
junction. We develop a self-consistent renormalization group
treatment of interacting phase-slips and their dual counterparts,
correlated cooper pair tunneling, beyond the dilute approximation.
This analysis leads to a very sharp transition with a critical
resistance of $R_Q$. The addition of the quasi-particles'
resistance at finite temperature leads to a quantitative agreement
with the experimental results. This self-consistent
renormalization group method should also be applicable to other
physical systems that can be mapped onto similar sine-Gordon
models, in the previously inaccessible intermediate-coupling
regime.
\end{abstract}

\end{frontmatter}

\section{Introduction}
\label{Introduction}

One of the most intriguing problems in low-dimensional
superconductivity is the understanding of the mechanism that
drives the superconductor-insulator transition (SIT). Experiments
conducted on quasi-one-dimensional (1D) systems have shown that
varying the resistivity and dimensions of thin metallic wires can
suppress superconductivity \cite{PXiongPRL1997,YOregPRL1999}, and
in certain cases lead to an insulating-like behavior
\cite{FSharifiPRL1993,ABezryadinNAT2000,CNLauPRL2001,ATBollinger2005,condmat}.

Particularly interesting are recent experiments conducted on short
MoGe nanowires \cite{ATBollinger2005} that explore the SIT tuned
by the wire's normal state resistance with a critical resistance
$R_c\approx R_Q$. Resistance measurements of the quasi-1D MoGe
nanowires reveal a strong temperature dependence that can be
fitted with a modified LAMH theory
\cite{JSLangerPR1967,DEMcCumberPRB1970} of thermally activated
phase-slips down to very low temperatures. However, for these
narrow wires it appears that the LAMH theory is valid only in a
narrow temperature window (see discussion in Sec. \ref{LAMH
validity}). Moreover, the LAMH analysis does not explain the
appearance of a critical value of $R_c\approx R_Q$.

The universal critical resistance may suggest that, at a
temperature much lower than the mean-field transition temperature,
$T\ll T_c $, the wire acts as a superconducting (SC) weak link
resembling a Josephson junction (JJ) connecting two SC leads.
Schmid~\cite{ASchmidPRL1983} and Chakravarty
\cite{SChakravartyPRL1982} showed that quantum phase-slip
fluctuations in such a JJ lead to a SIT as a function of the
junction's shunt resistance, $R_\mathrm{s}$, with a critical
resistance of $R_Q$, and that the resistance across the junction
obeys the power law $R(T)\propto
T^{2\left(\frac{R_Q}{R_\mathrm{s}} -1\right)}$. The theory was
later extended to JJ arrays and SC wires
\cite{HPBuchlerPRL2004,GRefael2005,GRefael2006}. Within these
theories, a similar power-law prevails. However, contrary to this
general prediction, Bollinger \textit{et
al.}~\cite{ATBollinger2005} observe that the resistance of the
MoGe wires exhibits a much stronger temperature dependence, even
close to the SIT.

In a previous work \cite{DMeidanPRL2007}, we have presented an
approach that captures both the critical resistance of $R_c\approx
R_Q$ at the SIT and the sharp decay of the resistance as a
function of temperature. We treat the SIT in nanowires as a
transition governed by quantum phase-slip (QPS) proliferation.
This picture alone, however, cannot account for the observed
strong temperature dependence of the resistance. We argued that
the key ingredient left out in previous works is the inclusion of
interactions between QPSs in such a finite-size wire, especially
when the phase-slip population is dense.

We treat these interactions in a mean-field type approximation:
when analyzing the behavior of a small segment of the wire, we
include in its effective shunt-resistance the resistance due to
phase-slips elsewhere in the wire. This scheme is motivated by
numerical analysis of a related problem, an interacting pair of
resistively-shunted JJs \cite{PWernerJSM2005}. This
self-consistent treatment primarily produces a sharp temperature
dependence of the resistance. In addition, we include the effects
of the Bogoliubov quasi-particles, which couple to the potential
gradient created by each
phase-slip~\cite{CommentBdGQuasiParticles}. Consequently, the
resistance obtained in the experiment can be fitted without
resorting to the LAMH theory beyond its limit of validity.

In this manuscript we generalize our previous results to the weak
Josephson coupling limit, where conductance through the wire
proceeds by Cooper pair tunneling. The use of a self-consistent
treatment of correlated Cooper pair tunneling events leads to a
similarly sharp SIT in the limit of highly resistive wires. This
self-consistent approximation of phase-slip interactions should be
applicable to similar multiple sine-Gordon models in the
theoretically challenging intermediate-coupling regime.

The remainder of the manuscript is organized as follows. In Sec.
\ref{LAMH validity} we discuss the validity of the LAMH theory for
the wires in Ref. \cite{ATBollinger2005}. In Sec. \ref{Low
resistive wires} we summarize our previous results on the role of
quantum phase-slip interactions in short wires, with details of
derivation of the microscopic model in Appendix \ref{Derivation of
Microscopic Action}. We generalize these results to highly
resistive wires in Sec. \ref{High resistive wires}, and present
our conclusion in Sec. \ref{conclusion}.

\section{On the validity of the theory of thermally activated phase-slips in short wires}
\label{LAMH validity}

\begin{figure}[h]
\begin{center}
\includegraphics[width=0.45\textwidth]{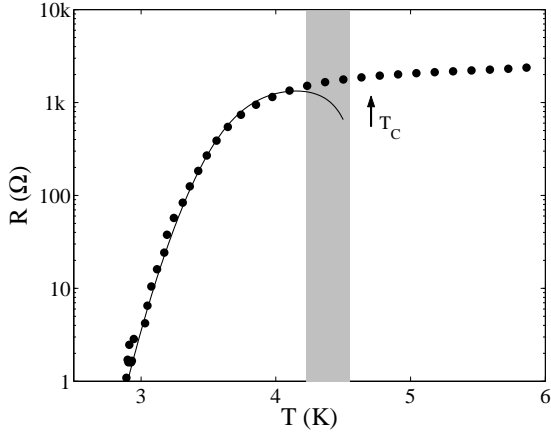}
\caption[0.6\textwidth]{Validity of the LAMH theory for the data
of wire No. 6 of Ref. \cite{ATBollinger2005} (see Table
\ref{param_table} for detailed parameters of the wire). The
fitting parameters used to fit the data are the coherence length,
$\xi=6.7$ nm, and the critical temperature, $T_c=4.7$ K, marked by
an arrow. These fitting parameters yield a temperature range $T^*
= 4.23 {\rm K} <T< T_G = 4.55$ K, for which the LAMH theory is expected
to be valid, shown in gray (see the text for a definition of $ T^*
$ and $T_G$).} \label{fig:validity}
\end{center}
\end{figure}

In an attempt to explain the strong temperature dependence of the
resistance of quasi-1D MoGe nanowires, Bollinger \textit{et
al.}~\cite{ATBollinger2005} have shown that the experimental
curves can be fitted with a modified LAMH theory
\cite{JSLangerPR1967,DEMcCumberPRB1970} of thermally activated
phase-slips, down to very low temperatures. Nevertheless,
estimations based on the parameters of the wires in Ref.
\cite{ATBollinger2005} (Table \ref{param_table}) suggest that most
of the temperature range in the experiment lies outside of the
region in which the theory is applicable. The theory of thermally
activated phase-slips is based on the time-dependent
Ginzburg-Landau description of a superconducting (SC) wire. This
description is valid at temperatures higher than the gap, and far
enough from $T_c$, such that fluctuation corrections are small,
$T^*<T<T_G$. Here $T^*$ is defined by $\Delta(T^*) = T^* $, with
$\Delta(T)$ the temperature dependent order parameter, and
$T_G=T_c(1- G\textit{i})$, with $G\textit{i} =\left[\frac{7
\zeta(3)}{4\pi^2}\frac{R_\xi}{R_Q}\right]^{\frac{2}{3}}$ the
Ginzburg-Levanyuk number for the quasi-1D wires, where $R_\xi=
R_{\mathrm{W}} \xi/L $ is the normal resistance of a section of the wire of
length $\xi$. For the wires in Ref. \cite{ATBollinger2005}, the
LAMH theory is valid only in a narrow temperature window as $T^*
\approx 0.9 T_c$, and estimates for $T_G $ range between $0.82
T_c$ and $0.97T_c$ (see Fig. \ref{fig:validity} and Table
\ref{param_table})~\cite{CommentLAMHValidityWiskers,RSNewbowerPRB1972}.
Moreover, the LAMH analysis does not explain the appearance of a
critical resistance $R_c\approx R_Q$.

This estimate suggests that most of the relevant temperature range
in the experiment conducted in Ref. \cite{ATBollinger2005} is in
the regime $T\ll \Delta(T)$, where quantum fluctuations become
increasingly important. While the universal critical resistance
supports the idea that the transition occurs due to quantum phase
slip proliferation in a narrow constriction, the naive extension
of the theory to SC wires \cite{HPBuchlerPRL2004,GRefael2005}
cannot account for the observed sharp drop of the resistance. In
the following section we will present a self-consistent treatment
of quantum phase-slip proliferation, that captures both the
critical resistance of $R_c\approx R_Q$ at the SIT and the sharp
decay of the resistance as a function of temperature.

\section{Low-resistance wires}
\label{Low resistive wires}

The microscopic action for a SC wire can be obtained from the BCS
Hamiltonian by a Hubbard--Stratonovich transformation followed by
an expansion around the saddle point (see Appendix \ref{Derivation
of Microscopic Action})\cite{CommentDirtyLimit}. In the limit of
low energy scales, $\omega, Dq^2~\ll~\Delta_0$, this yields
\cite{DSGolubevPRB2001}:
\begin{eqnarray}\label{effective_action}
\nonumber S &=& N_0A\Delta_0^2\int_0^L dx\int_0^{1/T}
d\tau\left\{\frac{\rho^2}{2}\left[\ln\left(\rho^2\right)-1\right]\right.\\
&&\left. +2\xi_0^2\rho^2\left[\phi'^2+\frac{\dot{\phi}^2
}{v_\phi^2}\right]+\xi_0^2\left[\rho'^2+\frac{\dot{\rho}^2}{v_{\rho}^2}
  \right]\right\},
\end{eqnarray}
where $L$ and $A$ are the wire's length and cross section,
respectively, $\xi_0^2=\pi D/8\Delta_0 $,
$v_\rho=\sqrt{(3\pi/2)D\Delta_0}$ the amplitude velocity,
$v_\phi=\sqrt{\pi D\Delta_0(2AV_cN_0+1)} $ the phase velocity,
$V_c$ the Fourier transform of the short-range Coulomb
interaction, $N_0$ the density of states, $D$ the electronic
diffusive constant in the normal state, and the SC order parameter
is parameterized as $\Delta = \Delta_0\rho e^{i\phi}$, with
$\Delta_0$ the mean-field solution. For the wires in
Ref.~\cite{ATBollinger2005}, $2AV_cN_0\propto N_\bot\sim 1000\gg 1
$, leading to $v_\rho\ll v_\phi\propto v_\rho\sqrt{N_\bot}$. Here
$N_\bot=p_F^2A /\pi^2$ is the number of 1D channels in the wire.

This action supports QPS excitations, which are characterized by
two distinct length scales: $v_\rho/\Delta_0\propto \xi\ll
\xi\sqrt{N_\bot}\propto v_\phi/\Delta_0$. For very long wires, $
\xi\ll \xi \sqrt{N_\bot}\ll L$, in the dilute phase-slip
approximation, this problem can be mapped onto the perturbative
limit of the $1+1$-dimensional sine-Gordon model \cite{HPBuchlerPRL2004}. In
the opposite limit of very short wires, $ L<\xi\ll \xi
\sqrt{N_\bot}$, the system resembles a JJ and can be mapped onto
the $0+1$-dimensional sine-Gordon model. However the wires in
Ref.~\cite{ATBollinger2005} appear to be in the intermediate
regime, $\xi \ll L\ll \xi \sqrt{N_\bot}$. Hence, while phase-slips
occur in different sections of the wire, they are
indistinguishable, as each creates a phase fluctuation that
spreads over distances larger than the wire itself.

Moreover, the wires in Ref. \cite{ATBollinger2005} have a sizable
bare fugacity. Using Eq. (\ref{effective_action}), one can
estimate the core action of a phase-slip of duration $\tau_0 =
1/\Delta_0$. Choosing the following  trial function for a phase
slip:
\begin{eqnarray}
\nonumber
  \phi &=& \arctan{\left(v_{\phi}\tau/x\right)} \\
  \rho&=&\min{[\sqrt{\left(x/x_0\right)^2+
  \left(\Delta_0\tau\right)^2},1]},
\end{eqnarray}
and identifying the part of the action [Eq.
(\ref{effective_action})], that  corresponds to $\rho \neq 1$ as
the core action, we minimize this expression with respect to the
phase-slip diameter, $x_0 $. Keeping only leading terms in
$N_\bot=\left(p_F^2A /\pi^2\right)$, this yields a core action of
$S_c =\pi/8\sqrt{1/3(1-\epsilon)}R_Q/R_\xi$
\cite{RSmithUnPublished}. Here $\epsilon = 1-T/T_c$ is the reduced
temperature and $R_\xi= R_{\mathrm{W}} \xi/L $ is the normal
resistance of a section of length $\xi$. Using this expression, the
measured values of $R_{\mathrm{W}}$ and $L$ (Table
\ref{param_table}), and the assumption that $\xi\approx 20$ nm,
the phase-slip fugacity is estimated to be $\zeta = e^{-S_c}\approx
0.05-0.45$ for the different wires. Consequently, in the critical
region, $R_{\mathrm{W}} \approx R_Q $, there is a dense population
of phase-slips that interact with one another; thus, the dilute
phase-slip approximation is no longer a proper description.

The flow equation for the fugacity of a phase-slip anywhere in the
wire is given by:
\begin{eqnarray}\label{RG_fugacity}
 \frac{d\zeta}{dl} &=& \left(1-
   \frac{R_Q}{R_\mathrm{s}}\right)\zeta,
\end{eqnarray}
where $dl = -d\ln\Lambda $, and $\Lambda $ is the running RG
scale. Eq.~(\ref{RG_fugacity}) treats the phase-slip as occurring
on an effective JJ, with $R_\mathrm{s}$ being the effective
shunting resistance of the entire wire. When $\zeta$ is small,
$R_\mathrm{s}$ will include only the effective impedance of the
leads. If $\zeta$ is not very small, we will need to include in
Eq.~(\ref{RG_fugacity}) additional terms of higher powers of
$\zeta$, which describe interactions between QPSs. To deal with a
finite $ \zeta$, we include the resistance due to other phase-slips
in the wire in the effective shunt resistance of the junction,
$R_\mathrm{s}$ \cite{DMeidanPRL2007}. This is akin to guessing the
form of a complete resummation of higher-order $\zeta$ terms in
Eq.~(\ref{RG_fugacity})~\cite{CommentNextOrderInRG,BulgadaevPLA1981}.

This treatment was successfully tested by numerical analysis of a
simpler analog of the system, an interacting pair of resistively
shunted JJs \cite{PWernerJSM2005}. In this work, Werner \emph{et.
al} studied the phase diagram and critical properties of the pair
of JJs using Monte Carlo simulations and renormalization
group calculations. The authors found that, in the region of the
intermediate coupling fixed point, there is a remarkable
resemblance in the critical behavior between the two-junction
system and a single junction. In order to explain this
resemblance, Werner \emph{et. al} suggest that the two-junction
system can be described by an approximate mean-field theory. In
the mean-field approximation, each junction at criticality behaves
as an independent junction and sees the other junction as an
effective resistor whose resistance is determined by phase-slip
events. This picture can account for the observed (and calculated)
properties of the two junction-system at the superconducting to
normal transition point.

The main physical intricacy of the self-consistent approach is the
determination of the effective shunt resistance,
$R_\mathrm{s}(\zeta)$, that governs the renormalization of the
phase-slip fugacity [Eq. (\ref{RG_fugacity})]. A phase-slip
produces time-varying phase gradients, and hence electrical
fields. These dissipate through two channels in parallel: the SC
channel - which has an effective resistance due to other phase
slips, $R_\mathrm{{ps}}$ - and the quasi-particles conduction
channel \cite{CommentBdGQuasiParticles}, which has resistance $R_\mathrm{{qp}} $. Once
the disturbance reaches the leads, it also dissipates through the
electro-dynamical modes of the large electrodes, whose real
impedance is parameterized by $R_\mathrm{{elec}}$.

For $T\ll~T_c $, the resistance of the quasi-particles,
$R_\mathrm{{qp}} $, can be approximated by
\begin{eqnarray}
R_\mathrm{{qp}} &=& \frac{m}{e^2\tau_{\mathrm{n}}
n_\mathrm{{qp}}}\frac{L}{A}= R_{\mathrm{n}}
\frac{n}{n_\mathrm{{qp}}}
   \approx R_{\mathrm{n}} \sqrt{\frac{T}{2\pi
   \Delta_0}}e^{\frac{\Delta_0}{T}}.
\end{eqnarray}

\begin{figure}[h]
\begin{center}
\includegraphics[width=0.5\textwidth]{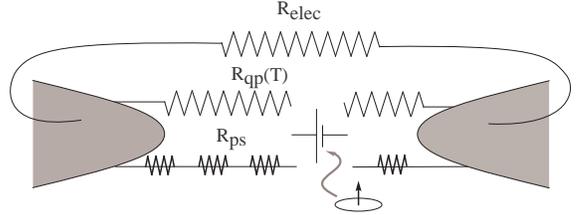}
\caption[0.5\textwidth]{ The effective electrical circuit for the
nanowire. A phase-slip produces time-varying phase gradients, and
hence electrical fields, that dissipate through the
quasi-particle conduction channel, which has resistance $R_\mathrm{{qp}} $, and
through the SC channel, that has an effective resistance due to
other phase-slips, $R_\mathrm{{ps}} $. Once the disturbance
reaches the edges, it also dissipates through the
electro-dynamical modes of the large electrodes, whose real
impedance is represented by $R_\mathrm{{elec}} $. }
\label{fig:circuit}
\end{center}
\end{figure}

Unfortunately, we lack a microscopic model for the impedance of
the electrodes, as this depends on the details of the system such
as the junction's shape and material. However, we expect that at
large scales, $T<\Lambda<\Delta $, the electrodes will act as a
transmission line to the electromagnetic waves generated by the
phase-slip. This transmission line is characterized by a real
impedance which we denote as $R_\mathrm{{elec}}$, and use as a
fitting parameter. Hence, the effective shunt resistance that
affects the renormalization of the phase-slip fugacity at
$T<\Lambda<\Delta $ [Eq. (\ref{RG_fugacity})] is
\begin{eqnarray}\label{effective_shunt_resistance}
    R_\mathrm{s}[\zeta(\Lambda)] &=&
    R_\mathrm{{elec}}+\left(\frac{1}{R_\mathrm{{ps}}[\zeta(\Lambda)]}+\frac{1}{R_\mathrm{{qp}}(T)}\right)^{-1}.
\end{eqnarray}
Fig. \ref{fig:circuit} shows the circuit we suggest describes the
system.

The resistance is measured in response to an applied DC current.
In this zero frequency limit, the electrodes act as a capacitor
connected in parallel to the wire. Therefore, the measured
resistance is the total wire resistance, unaffected by the
environment, which is cut off from the wire:
\begin{eqnarray}\label{total_DC_resistance}
  R_\mathrm{{tot}}(T)&=& \left(1/R_\mathrm{{ps}}[\zeta(T)]+1/R_\mathrm{{qp}}(T)\right)^{-1}.
\end{eqnarray}

The occurrence of a phase-slip causes a resistance in the
otherwise SC wire through the relation $R_\mathrm{{ps}}\propto
(L/\xi) \zeta^2 $. Using this relation and Eq.
(\ref{RG_fugacity}), we can write an RG equation for the
dimensionless resistance
 \begin{eqnarray}\label{RG_resistance}
 \nonumber
   \frac{d\zeta^2}{dl} &=& 2\zeta\frac{d\zeta}{dl} =2 \left(1- \frac{R_Q}{R_\mathrm{{s}}(\zeta)}\right)\zeta^2 \\
   \Rightarrow\frac{d\left(R_\mathrm{{ps}}/R_Q\right)}{dl} &=& 2 \left(1-
   \frac{R_Q}{R_\mathrm{{s}}(R_\mathrm{{ps}})}\right)\left(\frac{R_\mathrm{{ps}}}{R_Q}\right).
 \end{eqnarray}
Integration of Eq.~(\ref{RG_resistance}), with the effective
resistance given in Eq.~(\ref{effective_shunt_resistance}), from the
ultraviolet (UV) cutoff $\Delta(T^*)=T^*$ to the infrared
cutoff $T$, yields~$R_\mathrm{{ps}}(T) $.

\begin{figure}[h]
\begin{center}
\includegraphics[width=0.5\textwidth]{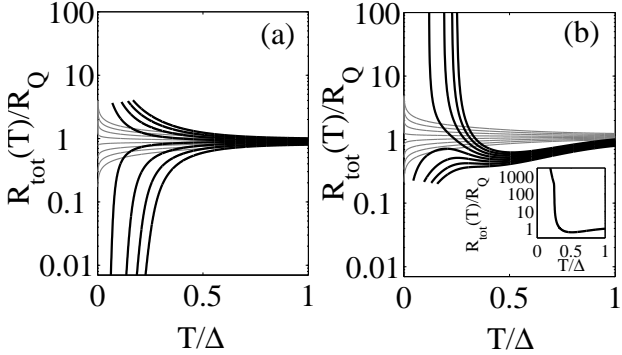}
\caption[0.5\textwidth]{The total wire resistance,
$R_\mathrm{{tot}}$, in units of $R_Q$ as a function of renormalized
temperature $T/\Delta $ for different $R^*/R_Q$, ranging from
$0.8$ (lower plot) to $1.01$ (upper plot) in increasing steps of
$0.03$. The environment resistance was taken to be
$R_\mathrm{{elec}}/R_Q= 0.1$. Gray traces compare the total wire
resistance $R_\mathrm{{tot}}(T)=\frac{R(T)
R_\mathrm{{qp}}(T)}{R(T)+R_\mathrm{{qp}}(T)} $ to a single
Josephson junction with $R(T) =
R^*\left(T/\Delta\right)^{2(R_Q/R_\mathrm{s}-1)}$, and constant
$R_s$.  \textbf{(a)} The total wire resistance,
$R_\mathrm{{tot}}(T)=\frac{R_\mathrm{{ps}}
R_\mathrm{{qp}}(T)}{R_\mathrm{{ps}}+R_\mathrm{{qp}}(T)} $, in the
limit of weak phase-slip fugacity calculated using Eqs.
(\ref{effective_shunt_resistance}) and (\ref{RG_resistance}). The
insulating plots are cutoff at temperature, $T_0 $, for which
$\zeta(T_0) = 1 $ [see the discussion prior and subsequent to Eq.
(\ref{effective_shunt_resistance_zeta})]. \textbf{(b)} The total
wire resistance, $R_\mathrm{{tot}}(T)=1/G_\mathrm{{tot}}(T) =
\left(G_\mathrm{{cp}}(T)+1/R_\mathrm{{qp}}(T)\right)^{-1}$, in the
weak Josephson coupling limit calculated using Eqs.
(\ref{dual_self_consistancy}) and (\ref{dual_effective shunt
resistance}). Here we have assumed the wire length is $L=5\xi $,
where $\xi$ is the coherence length. The initial drop in the
resistance of the high-resistance wires is a manifestation of the
difference between the total wire resistance and the effective
shunt resistance, in the presence of local quasi-particle
relaxation, in the weak coupling limit. As the ratio between the
wire length and the coherence length increases, the initial drop
in the resistance becomes more pronounced. The low-resistance plots
are cutoff at $T_0 $, for which $J(T_0) = 1 $. Inset: The
resistance of the wire calculated in the self consistent
approximation, Eq. (\ref{dual_self_consistancy}), shows an initial
sharp increase as the temperature is lowered, followed by a
moderate increase. This moderation of the diverging resistance is
due to the finite density of quasi-particles, present at finite
temperature.} \label{fig:high_vs_low_resistive}
\end{center}
\end{figure}

The wire's DC resistance Eq. (\ref{total_DC_resistance}),
calculated using Eqs. (\ref{effective_shunt_resistance}) and (\ref{RG_resistance}), is plotted in
Fig~\ref{fig:high_vs_low_resistive} (a) as a function of
temperature for different $R^*$. Here $R^*\equiv
R_\mathrm{{tot}}(T^*)$ is the normal state resistance of the wire,
at the UV cutoff $\Delta(T^*)=T^*$. We assume that the wires are
thin enough such that the mean-field transition from normal to SC
is wide, and $R^*\approx R_\mathrm{{tot}}(T_c^+)=R_{\mathrm{n}} $.
For simplicity, we have assumed throughout our calculation that
$\Delta(T)\approx\Delta_0 $. This assumption holds in the low-temperature regime, $T\ll \Delta$, where the theory of QPSs, based
on the effective action Eq. (\ref{effective_action}), is expected
to be valid. The resistance of the environment is taken to be
$R_\mathrm{{elec}}=0.1 R_Q$. In practice, $\Delta$, $R^*$ and
$R_\mathrm{{elec}}$ can be used as fitting parameters. Moreover,
$R^* $ can be determined independently as the resistance measured
below the drop that indicates passing through $T_c$ of the SC
films (see Ref.~\cite{ATBollinger2005}).

Eq. (\ref{RG_resistance}) is also applicable to a wire with
$R^*>R_Q$. In this limit the fugacity increases in the
renormalization process, and Eq. (\ref{RG_fugacity}) is no longer
valid for $\zeta\gtrsim 1 $. For wires with $R^*>R_Q $, we
overestimate the phase-slip fugacity as $\zeta(T^*)=0.5 $ and
integrate Eq.(\ref{RG_fugacity}), with
\begin{eqnarray}\label{effective_shunt_resistance_zeta}
  R_\mathrm{s}[\zeta(\Lambda)]
&=&R_\mathrm{{elec}}+\left(\frac{1}{y_\zeta(R^*)\zeta(\Lambda)^2}+\frac{1}{R_\mathrm{{qp}}(T)}\right)^{-1}\!\!\!\!\!\!\!\!,
\end{eqnarray}
from the UV cutoff, $\Delta(T^*)=T^*$, down to the
temperature $T_0$ for which $\zeta(T_0)=1$. Here the
proportionality constant $y_\zeta(R^*)$ is set by the initial
condition $R^* =
\left(1/[y_\zeta(R^*)\zeta(T^*)^2]+1/R_\mathrm{{qp}}(T)\right)^{-1}$,
with $\zeta(T^*)=0.5 $. The results are shown in
Fig.~\ref{fig:high_vs_low_resistive}~(a). Overestimating
$\zeta(T^*)$ gives an upper bound on $T_0$, where Eq.
(\ref{RG_fugacity}) is no longer applicable.
Fig.~\ref{fig:high_vs_low_resistive}~(a) shows that the transition
between SC and insulating wires occurs for a critical resistance
$R_c\approx R_Q$. However, in contrast to the standard Josephson
junction theory (gray curves in
Fig.~\ref{fig:high_vs_low_resistive}~(a)), the transition is much
sharper (notice the logarithmic scale).

\begin{table}[ht]
\begin{center}
\begin{tabular}[width=0.5\textwidth]{| c | r  r @{.} l | r @{.} l  r @{.} l  r @{.} l |}
\hline
 &\multicolumn{3}{|c|}{Measured
Values}&\multicolumn{6}{|c|}{Fitting Values}\\
\hline
Curve & $L (nm)$ &\multicolumn{2}{c|}{$R_{\mathrm{W}} (k\Omega)$} &\multicolumn{2}{c}{$R^* (k\Omega)$} &\multicolumn{2}{c}{$\Delta(K)$} &\multicolumn{2}{c|}{$R_\mathrm{{elec}} (k\Omega)$}  \\
\hline
$1$ & 177\phantom{0} & \phantom{00}5&46 & \phantom{00}4&2 &  \phantom{0}2&5 & \phantom{00}1&2\\
$2$ & 43\phantom{0} & 3&62 & 2&62 & 2&35 & 1&25\\
$3$ & 63\phantom{0} & 2&78 & 2&13 & 3&07 & 0&66\\
$4$ & 93\phantom{0} & 3&59 & 2&89 & 3&85 & 0&55\\
$5$ & 187\phantom{0} & 4&29 & 4&5 & 6&55 & 0&31\\
$6$ & 99\phantom{0} & 2&39 & 2&09 & 4&84 & 0&4\\
\hline
\end{tabular}
\end{center}
\caption[width=0.5\textwidth]{Summary of nanowire parameters, and
the parameters used to fit the experimental data; $L$ is the
length of the wire determined from SEM images, $ R_{\mathrm{W}}$
is the wire's normal state resistance, taken as the resistance
measured below the film transition. Fitting parameters:
$R^*=R_{\mathrm{tot}}(T = \Delta) $ is the wire's resistance at
the UV cutoff, $\Delta $ the SC order parameter, and $
R_\mathrm{{elec}}$ the impedance of the electrodes at
$T<\Lambda<\Delta $.} \label{param_table}
\end{table}

\begin{figure}[h]
\begin{center}
\includegraphics[width=0.5\textwidth]{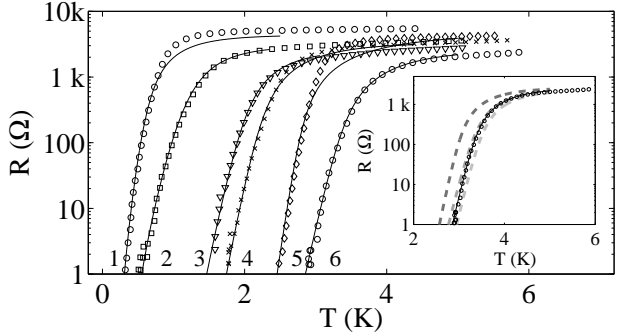}
\caption[0.5\textwidth]{Comparison between the theory (solid line)
and the experimental data \cite{ATBollinger2005}. Details of the
wires, and fitting parameters are summarized in Table
\ref{param_table}. Note that the estimated values for $R^*$ are in
good agreement with the measured values of the resistance of the
wire at $T=\Delta $. The discrepancies between theory and
experiment occur for the longer wires (curves $1 $ and $ 5$). In
these wires the condition $v_\phi/\Delta_0 \gg L$ might not hold,
and one should consider additional renormalization at higher
scales. Inset: The effect of an adjustment of the fit parameters,
in roughly $10\%$ (from dark gray to light gray): $\Delta = 4.356
K$, $R_\mathrm{{elec}}=440 \Omega$ and $R^* = 1881 \Omega $.}
\label{fig:fit}
\end{center}
\end{figure}

A comparison between the theoretical curves and the experimental
data taken from Ref. \cite{ATBollinger2005} is shown in Fig.~
\ref{fig:fit}. The curves were calculated by fitting $R^* $,
$\Delta$, and $R_\mathrm{{elec}}$. Since the theory of QPSs is
expected to be valid at $T\ll \Delta $, deviations from the
theoretical curves at high temperature are reasonable. In general,
as $\Delta$ is proportional to $T_c $, increasing $\Delta$ shifts
the sharp decay of the resistance to high temperatures. Both $R^*$
and $R_\mathrm{{elec}}$ affect the high temperature resistance,
whereas $R^*$ and $\Delta $ control the width of the transition.

We have made an attempt to fit the data corresponding to the
insulating wires of Ref. \cite{ATBollinger2005}. As the insulating
wires are thinner, we expect a strong suppression of $T_c$
\cite{YOregPRL1999}, which sets the scale for the high temperature
cutoff $T\ll\Delta(T)$. Moreover, in these wires, the bare
fugacity is estimated by $\zeta = e^{-S_c}\approx 0.32-0.84$,
which results in a relatively high $T_0$, for which
$\zeta(T_0)=1$. Consequently, the theory of QPSs with
$\zeta\lesssim 1$ is valid in a narrow range of temperatures.
While we manage to fit the experimental curves in this regime of
parameters, we do not present the results as our fits cover a
range of $\sim 10 $ data points, with three fitting parameters.

\section{High-resistance wires}
\label{High resistive wires}

To better describe the insulating wires, we turn to the weak
Josephson coupling limit where phase coherence across the wire is
lost, and conductance proceeds by means of Cooper pair tunneling
events across regions of fluctuating order parameter amplitude.
The flow equation for the Josephson coupling anywhere in the wire
is given by:
\begin{eqnarray}\label{RG_Josephson}
  \frac{d J}{dl} &=& \left(1-
   \frac{R_\mathrm{s}}{R_Q}\right)J.
\end{eqnarray}
Once more we expect that the Josephson tunneling between two such
sections will be effected by all other tunneling events in the
wire, via capacitive coupling.

\begin{figure}[h]
\begin{center}
\includegraphics[width=0.5\textwidth]{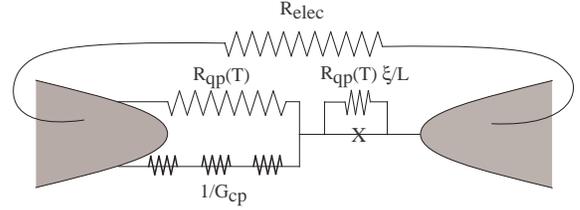}
\caption[0.5\textwidth]{ The effective electrical circuit for the
nanowire in the weak coupling limit. When a Cooper pair tunnels a
distance $\xi$, a dipole of strength $2e\xi$ forms. It can relax
either by locally fusing back through the quasi-particle channel,
i.e, through a resistance $R_{\mathrm{{qp}}}(T) \xi/L$, or by
increasing the separation between the tunneling Cooper pair and
the hole it leaves behind, until they leave the system through the
electrodes. The advancing Cooper pair can relax by means of
successive tunneling events with resistance $1/G_{cp} $, or by
exciting quasi-particles along the wire, with resistance $R_{qp}(T)$. Once more,
as the excitation reaches the edges, it dissipates through the
electro-dynamical modes of the large electrodes, whose real
impedance is represented by $R_\mathrm{{elec}} $. }
\label{fig:circuit_weak_coupling}
\end{center}
\end{figure}

The pair tunneling event leads to a finite conductance through the
insulating wire, $G_\mathrm{{cp}}\propto J^2 $. To use this
relation and Eq. (\ref{RG_Josephson}), we need to consider what
the effective shunting resistance of an individual Josephson
junction (a segment of length $\xi$) is. When a Cooper pair
tunnels a distance $\xi$, a dipole of strength $2e\xi$ forms. It
can relax either by locally fusing back through the quasi-particle
channel, i.e, through a resistance $R_{\mathrm{{qp}}}(T) \xi/L$,
or by the tunneling Cooper pair, and the hole it leaves behind
increasing their separation, until they leave the system through
the electrodes. The latter process implies the $2e$ charge goes
through a resistance
$R_\mathrm{elec}+\left(G_\mathrm{{cp}}[J(\Lambda)]+\frac{1}{R_\mathrm{{qp}}(T)}\right)^{-1}$.
These two channels appear in parallel, see Fig.
\ref{fig:circuit_weak_coupling}. Therefore the full shunting
resistance is:
\begin{eqnarray}\label{dual_effective shunt resistance}
 \nonumber
  &&R_{\mathrm{s}}[J(\Lambda)]^{-1}=\\
  &&\frac{1}{
R_\mathrm{elec}+\left(\frac{R_\mathrm{{qp}}(T)}{R_\mathrm{{qp}}(T)G_\mathrm{{cp}}[J(\Lambda)]+1}\right)}+\frac{1}{R_\mathrm{{qp}}(T)\frac{\xi}{L}}.
\end{eqnarray}

Thus the RG equation for the conductance, in units of
$G_Q=4e^2/h$, is
\begin{eqnarray}\label{dual_self_consistancy}
   \frac{d}{dl}\left(\frac{G_\mathrm{{cp}}}{G_Q}\right) &=&
  2 \left[1-\frac{R_{\mathrm{s}}(\Lambda)}{R_Q}
   \right]\left(\frac{G_\mathrm{{cp}}}{G_Q}\right),
 \end{eqnarray}
where $R_\mathrm{s}$ is given by Eq. (\ref{dual_effective shunt
resistance}). The total resistance of the wire,
$R_\mathrm{{tot}}(T)=1/G_\mathrm{{tot}}(T) =
\left(G_\mathrm{{cp}}[J(T)]+1/R_\mathrm{{qp}}(T)\right)^{-1}$,
calculated using Eqs. (\ref{dual_effective shunt resistance}) and (\ref{dual_self_consistancy}), is plotted in
Fig~\ref{fig:high_vs_low_resistive}~(b) as a function of
temperature for different $R^*$. As mentioned above, $R^*\equiv
R_\mathrm{{tot}}(T^*)\approx R_{\mathrm{n}}$ is the normal state
resistance of the wire, at the UV cutoff $\Delta(T^*)=T^*$, which
can be determined from resistance measurements at a temperature
below the SC transition of the 2D films \cite{ATBollinger2005}.
The order parameter is assumed to be constant throughout the
calculation and the resistance of the environment is taken to be
$R_\mathrm{{elec}}=0.1 R_Q$. In Fig.
~\ref{fig:high_vs_low_resistive}~(b) we have assumed $L = 5\xi $.
In Ref. \cite{ATBollinger2005} the high-resistance wires are
typically thinner, and are therefore expected to have a strong
suppression of $T_c$ \cite{YOregPRL1999}. This will lead to a
relatively large coherence length, and a smaller ratio $L/\xi $.

The initial drop in the resistance of the high-resistance wires is
a manifestation of the difference between the total wire
resistance and the effective shunt resistance, in the weak
coupling limit. The local quasi-particle relaxation reduces the
shunt resistance below $R_Q $ at $T\sim \Delta(T) $, causing an
initial drop in the resistance. At lower temperature, the density
of quasi-particles decreases exponentially, and
$R_{\mathrm{qp}}\xi/L$ ceases to be smaller than $R_Q$. This leads
to a decrease in the number of tunneling events causing the total
resistance to increase. As the ratio between the length of the wire and
the coherence length increases, the initial shunt resistance is
smaller and the initial drop in the resistance becomes more
pronounced.

The resistance of the wire, calculated in the self-consistent
approximation, Eq. (\ref{dual_self_consistancy}), shows an initial
sharp increase as the temperature is lowered, followed by a
moderate increase. As the pair tunneling drops rapidly to zero at
a finite temperature, conductance persists due to the finite
density of quasi particles. This residual conductance appears as a
moderation of the diverging resistance, as shown in
Fig~\ref{fig:high_vs_low_resistive}~(b). Conversely, upon
neglecting the interactions between different sections of the
wire, the conductance due to independent Josephson tunneling shows
a power law decay as a function of decreasing temperature,
$G_\mathrm{{cp}}(T) =
1/R^*\left(T/\Delta\right)^{-2(R_Q/R_\mathrm{s}-1)}$. Such a power
law behavior leads to a finite conductance at finite temperature,
which shunts the highly resistive contribution of the
quasi-particles.

Eq. (\ref{dual_self_consistancy}) is also applicable to a wire
with $R^*<R_Q$. In this limit the pair tunneling increases in the
renormalization process, and Eq. (\ref{dual_self_consistancy}) is
no longer valid for $J\gtrsim 1 $. We estimate bare Josephson
coupling as $J(T^*)=0.5 $ and integrate Eq.~(\ref{RG_Josephson})
with
\begin{eqnarray}
  R_{\mathrm{s}}(\Lambda) &=&
  R_\mathrm{elec}+\left(y_J(R^*)J(\Lambda)^2+\frac{1}{R_\mathrm{{qp}}(T)}\right)^{-1},
\end{eqnarray}
from the UV cutoff, $\Delta(T^*)=T^*$, down to the
temperature $T_0$ for which $J(T_0)=1$. Once more, the
proportionality constant is set by the initial condition
$R^*=\left[y_J(R^*)J(T^*)^2 +1/R_\mathrm{{qp}}(T)\right]^{-1}$.
The results are shown in Fig.~\ref{fig:high_vs_low_resistive} (b).

When we use the resistive wire theory, we must consider the
following caveat: If the Josephson coupling between segments of
the wire is only perturbative, then the phase of the order
parameter fluctuates locally as a function of time. If the
frequency scale of this fluctuation is comparable to the
order-parameter magnitude, then there should also be low lying sub-gap density of states in the wire, that we don't take into
account. The theory in this section assumes that the phase
fluctuations are sufficiently slow such that the sub-gap states
can be ignored.

\section{Conclusion}
\label{conclusion}

In conclusion, we studied the effect of interactions between QPSs
in short SC wires, beyond the dilute phase-slip approximation. Our
analysis shows that treating these interactions in a self-consistent manner produces a sharp superconductor-insulator transition with a
critical resistance $R_c\approx R_Q$, in agreement with recent
experiments \cite{ATBollinger2005}. Moreover, we have shown that
adding the resistance of the BdG quasi-particles leads to a
quantitative agreement with the experimental curves. In the dual
weak Josephson coupling limit, this self-consistent RG treatment
produces a similar sharp insulator-superconductor transition. The
sharp drop in the conductance of the insulating wires in this
limit is shown to be accompanied by a residual conductance due to
the density of quasi-particles at this finite temperature. Our
method should be applicable to a wider range of physical problems
which involve the proliferation of topological defects with a
sizable bare fugacity. In particular, it could be applied to the
study of a Luttinger liquid with an extended
impurity~\cite{CLKanePRB1992}.

We thank E.~Demler, and P. Werner. Special thanks to
A.~Bezryadin for making his data available to us. This study was
supported by a DIP grant and by an ISF grant.

\appendix
\section{Derivation of microscopic action}
\label{Derivation of Microscopic Action}

Consider a system of electrons in the diffusive limit that
interact via Coulomb repulsion and the phonon mediated BCS
interaction. This is described by the Hamiltonian
$H=H_0+H_{int}+H_{BCS} $:
\small
\begin{eqnarray}
 \nonumber
  H_0 &=& \sum_\sigma\int d^3r\psi_\sigma^\dag(r)\left(-\frac{\nabla_r^2}{2m}+\sum_i u(r-r_i)\right)\psi_\sigma^\dag(r)\\
  \nonumber
  H_{int} &=& \sum_{\sigma,\sigma '}\int d^3r d^3r' \psi_\sigma^\dag(r)\psi_\sigma(r)\frac{e^2}{|r-r'|}\psi_{\sigma '}^\dag(r')\psi_{\sigma '}(r')\\
  H_{BCS} &=& -\lambda\sum_{\sigma}\int d^3r
  \psi_\sigma^\dag(r)\psi_\sigma(r)\psi_{\bar{\sigma}}^\dag(r)\psi_{\bar{\sigma}}(r),
\end{eqnarray}
\normalsize
where $\psi_\sigma^\dag(r)$ and $\psi_\sigma(r) $ are electron
creation and annihilation operators, the Fourier transform of the
Coulomb interaction in 1D $V_c(q)\sim \log{q} $ is taken to be
constant, and $u(r-r_i) $ is the impurity potential at point $r$
due to an impurity at point $r_i$. We assume that the impurity
potential is $\delta$ correlated, $\langle u(r)\rangle =0$, and
$\langle u(r)u(r')\rangle =\frac{1}{2\pi N_0\tau }\delta(r-r')$,
where $ \tau$ is the impurity scattering time, and $N_0$ is the 3D density of
states. Following Ref. \cite{UEckernJLTP1988}, we apply a
Hubbard--Stratonovich transformation in order to rewrite the BCS
interaction and the Coulomb interaction. The partition function
becomes
\begin{eqnarray}
\nonumber
  Z &=& \int D\Delta D\Delta^* D\rho D\psi D\psi^\dag e^{-S},
\end{eqnarray}
where
\begin{eqnarray}
\nonumber
  S &=& S_0+ \int dx\frac{|\Delta(x)|^2}{\lambda}\\
\nonumber
  &+& \int dx \left\{\Delta(x)^*\psi_{\downarrow}(x)\psi_{\uparrow}(x)+\Delta(x)\psi_{\uparrow}^\dag(x)\psi_{\downarrow}^\dag(x)\right\}\\
\nonumber
  &+&\frac{1}{2}\int dxdx' \rho(x)V_c^{-1}(x-x')\rho(x') \\
  &+&i\int dx \rho(x)\left\{\psi_{\uparrow}^\dag(x)\psi_{\uparrow}(x)
   +\psi_{\downarrow}^\dag(x)\psi_{\downarrow}(x)\right\},
\end{eqnarray}
and $x\equiv(\textbf{r},\tau)$.
Integrating over the fermionic
fields, the effective action in Nambu-Gorkov spinor notation reads
\begin{eqnarray}\label{effective_action_tr_log}
\nonumber
   S &=&\int dx\lambda^{-1}|\Delta(x)|^2  -Tr\ln{ G}^{-1}\\
   &&+\frac{1}{2}\int dxdx'
   \rho(x)V_c^{-1}(x-x')\rho(x')
\end{eqnarray}
with
\begin{eqnarray}
    G^{-1} &=& \left(%
\begin{array}{cc}
  \partial_\tau-\xi+i\rho & \Delta \\
  \Delta^* & \partial_\tau+\xi-i\rho \\
\end{array}%
\right),
\end{eqnarray}
and $\xi=-\nabla^2/2m-\mu $. One can treat the presence of nonmagnetic impurities by including a self-energy diagram that
describes the dressing of the electron line by impurities. As a
result, the frequencies and the order parameter are replaced by
$\tilde{\omega}=\eta\omega $ and $\tilde{\Delta}=\eta\Delta $,
with $\eta=\left(1+1/(2\tau \sqrt{\omega^2+\Delta_0^2}) \right)$
\cite{Kopnin2001}. Moreover, when calculating polarization bubbles
(see section \ref{fluctuations}), one must sum over the impurity
ladder (namely impurity lines connecting the two Green's functions
in the polarization bubble).

\subsection{Uniform order parameter}

In the case of a uniform order parameter, the action in Eq.
(\ref{effective_action_tr_log}) can be greatly simplified. In this
limit, the order parameter $\Delta$ can be chosen to be real and
the action reduces to \cite{UEckernPRB1984}
\begin{eqnarray}\label{action_homogenuos}
\nonumber
   S[\Delta] &=&\int dx\lambda^{-1}|\Delta(x)|^2 \\
   &&-
    T\sum_\omega\nolimits'\int
    \frac{d^3p}{(2\pi)^3}\ln{\left(\tilde{\omega}^2+\xi_p^2+\tilde{\Delta}^2\right)}.
\end{eqnarray}
A variation of the action in Eq. (\ref{effective_action_tr_log})
with respect to $\Delta^* $ yields the BCS gap-equation
\begin{eqnarray}
  \frac{\Delta_0 }{\lambda}&=&   \pi N_0T\sum_\omega\nolimits '
  \frac{\Delta_0}{\sqrt{\omega^2+\Delta_0^2}},
\end{eqnarray}
where the sum $\sum_\omega\nolimits' $ indicates that the
frequencies are cut off at the Debye frequency, $\omega_D $.
In~order to evaluate the sum in Eq. (\ref{action_homogenuos}), we
replace $S[\Delta]$ by
\begin{eqnarray}
  S[\Delta] - S[0] &=& \int_0^\Delta \frac{\partial S[\Delta]}{\partial
  \Delta},
 \end{eqnarray}
where
\begin{eqnarray}
\nonumber
    \frac{\partial S[\Delta]}{\partial\Delta} &=&
    \frac{2\Delta}{\lambda}-2\tilde{\Delta} T\sum_\omega\nolimits'\int
    d^3p\frac{1}{\left(\tilde{\omega}^2+\xi_p^2+\tilde{\Delta}^2\right)}\\
    &=& \frac{2\Delta}{\lambda}-2\pi N_0T\sum_\omega\nolimits '
  \frac{\Delta}{\sqrt{\omega^2+\Delta^2}}.
\end{eqnarray}
We have substracted $S[0]$ to avoid divergences. As $S[0]$ is
independent of $\Delta $, this choice does not affect our final
results. Integrating with respect to $\Delta$ results in the
following action
\begin{eqnarray}
\nonumber
  S[\Delta]-S[0] &=& \frac{\Delta^2}{\lambda}-4\pi N_0
    T\sum_{\omega=0}^{\omega_D}\left(\sqrt{\omega^2+\Delta^2}-\omega\right)\\
    &\approx&\frac{\Delta^2}{\lambda}-N_0\Delta^2\left[\frac{1}{2}+\ln{\left(\frac{2\omega_D}{\Delta}\right)}\right],
\end{eqnarray}
which can be written as
\begin{eqnarray}\label{order_parameter_magnitude}
     S[\Delta]-S[0]
     &=&\frac{N_0\Delta^2}{2}\left[\ln\left({\frac{\Delta^2}{\Delta_0^2}}\right)-1\right].
\end{eqnarray}

\subsection{Fluctuations around the mean-field
solution}\label{fluctuations}

The action describing phase and amplitude fluctuations is obtained
by expanding Eq. (\ref{effective_action_tr_log}) around the mean-field saddle point solution $\Delta = \Delta_0$. Dividing the
fluctuations into real and imaginary parts, $\delta \Delta =
\Delta_L+i \Delta_T$, which are connected to amplitude and phase
variations, the fluctuations around the mean field are given by
\begin{eqnarray}
\nonumber
   S &=&\int dx\frac{|\Delta_0
   |^2}{\lambda}+\frac{\Delta_L^2+\Delta_T^2}{\lambda}\\
\nonumber
   &&+\frac{1}{2}\int dxdx' \rho(x)V_c^{-1}(x-x')\rho(x')-Tr\ln{ G}_0^{-1}\\
   &&-Tr\sum_{n=2}^\infty \frac{(-1)^{n+1}}{n}\left({G}_0\delta{G}^{-1}\right)^n.
\end{eqnarray}
Here
\begin{eqnarray}
\nonumber
  {G}_0 &=& \frac{1}{\tilde{\omega}^2+\xi^2+\tilde{\Delta}_0^2}\left(%
\begin{array}{cc}
  -i\tilde{\omega}-\xi & \tilde{\Delta}_0 \\
  \tilde{\Delta}_0^* & -i\tilde{\omega}+\xi \\
\end{array}%
 \right)\\
  \delta{G}^{-1}&=&
\left(   \begin{array}{cc}
  i\rho & \Delta_L+i  \Delta_T \\
  \Delta_L-i\Delta_T & -i\rho\\
\end{array}%
 \right).
\end{eqnarray}
Keeping only leading terms in $\rho $, $\Delta_L $ and $\Delta_T$,
we find that the partition function can be written as
\begin{eqnarray}
  Z &=& \int D\Delta_L D\Delta_T\ D\rho e^{-S_{eff}\left[\Delta_L,\Delta_T,\rho\right]},
\end{eqnarray}
with the effective action
\begin{eqnarray}
  S_{\mathrm{eff}} &=& -\int{\,\mathchar'26\mkern-9mu{\!d}} q \left(%
\begin{array}{ccc}
  \Delta_L & \Delta_T & \rho \\
\end{array}%
\right)_q V^{-1}(q) \left(\!\!%
\begin{array}{c}
  \Delta_L \\
  \Delta_T \\
  \rho \\
\end{array}%
\!\!\right)_{-q}\!\!\!\!\!\!,
\end{eqnarray}
where we have introduced a shorthand notation
$\int{\,\mathchar'26\mkern-9mu{\!d}} q = A T\sum_{\omega_n} \int
dq/ 2\pi$ and $A$ is the wire's cross section. The screened potentials
are given by \cite{RASmithPRB1995}
\begin{eqnarray}
\nonumber
  && V^{-1}(q) =\\
   && \left(\!\!\!%
\begin{array}{ccc}
  -\frac{1}{\lambda}+\Pi_{\Delta_L\Delta_L}\!\! & 0 \!\!& 0 \\
  0 \!\!&  -\frac{1}{\lambda}+\Pi_{\Delta_T\Delta_T}\!\! & \Pi_{\Delta_T\rho }\\
  0\!\! & -\Pi_{\Delta_T\rho }\!\! & \frac{1}{2A V_c}+\Pi_{\rho\rho}
  \\
\end{array}%
\!\!\!\right).
\end{eqnarray}
In the above, $\Pi_{\alpha\beta}$ is the polarization bubble obtained by
integrating out the electronic degrees of freedom, with the
vertices $\alpha$ and $\beta$ corresponding to the incoming and
outgoing bosonic fields.

In the dirty limit $ql,\Omega \tau \ll 1$, where $q $ and $
\Omega$ are the transferred momentum and frequency, respectively,
and $l $ and $ \tau$ are the mean free path and impurity
scattering time, we find
\begin{eqnarray}\label{polarization_bubbles}
\nonumber
  \Pi_{\Delta_L\Delta_L}(q,\Omega) &=& \pi N_0 T\sum_{\omega}\left\{\left[1 +\frac{\omega\omega'-\Delta_0^2}{WW'}\right]\right.\\
 \nonumber
 &\times& \left.\frac{1}{W+W'+Dq^2} \right\}\\
\nonumber
  \Pi_{\Delta_T\Delta_T}(q,\Omega) &=& \pi N_0 T\sum_{\omega}\left\{\left[1 +\frac{\omega\omega'+\Delta_0^2}{WW'}\right]\right.\\
 \nonumber
 &\times&\left. \frac{1}{W+W'+Dq^2} \right\}\\
\nonumber
  \Pi_{\rho\rho}(q,\Omega) &=& N_0- \pi N_0 T\sum_{\omega}\left\{\left[1 -\frac{\omega\omega'+\Delta_0^2}{WW'}\right]\right.\\
   \nonumber
 &\times&\left. \frac{1}{W+W'+Dq^2} \right\}\\
  \nonumber
 \Pi_{\Delta_T\rho}(q,\Omega) &=& -\pi N_0
T\sum_{\omega}\frac{\Delta_0
  \Omega}{WW'}\frac{1}{W+W'+Dq^2}\\
  &=& -\Pi_{\rho\Delta_T}(q,\Omega).
\end{eqnarray}
Here $ \omega' = \omega+\Omega$, $W = \sqrt{\omega^2+\Delta_0^2}
$, $W' = \sqrt{\omega'^2+\Delta_0^2} $ and $D =\tau v_F^2/3 $ is
    the diffusion constant. In the low temperature limit,
for $q \ll \xi\sim\xi_0 $ and $\Omega\ll \Delta_0 $, where $\xi_0
$ is the zero temperature coherence length, these may be
approximated by
\begin{eqnarray}\label{screened_interactions}
\nonumber
 && \Pi_{\Delta_L\Delta_L}(q,\Omega)-\lambda^{-1} \approx -N_0\left(1+\frac{\Omega^2}{12 \Delta_0^2}+\frac{\pi}{8}\frac{Dq^2}{\Delta_0}\right)\\
\nonumber
&&  \Pi_{\Delta_T\Delta_T}(q,\Omega)-\lambda^{-1} \approx -N_0\left(\frac{\Omega^2}{4 \Delta_0^2}+\frac{\pi}{4}\frac{Dq^2}{\Delta_0}\right)\\
\nonumber
 && \Pi_{\rho\rho}(q,\Omega) \approx N_0\left(1-\frac{1}{6}\frac{\Omega^2}{\Delta_0^2}\right) \\
 && \Pi_{\Delta_T\rho}(q,\Omega) \approx -\frac{N_0\Omega}{2
  \Delta_0}\left(1-\frac{1}{6}\frac{\Omega^2}{\Delta_0^2}-\frac{\pi}{8}\frac{Dq^2}{\Delta_0}\right)\!\!.
\end{eqnarray}

Note that close to $T_c$, namely in the limit  $\Delta\ll \Omega,
Dq^2\ll T $, the polarization bubble
$\Pi_{\Delta_L\Delta_L}=\Pi_{\Delta_T\Delta_T}\approx
-N_0\left(\frac{\pi}{8}\frac{|\Omega
|+Dq^2}{T}-\ln\left(\frac{1.13\omega_D}{T}\right)\right)$, which
reproduces the time-dependent Ginzburg-Landau (TDGL). The term
$\sim |\Omega |$ describes dissipation. Its emergence is a result
of the fact that the polarization bubbles describe the response of
the electronic system, in equilibrium. Hence, we have assumed the
existence of a relaxation mechanism that allows the electrons to
return to equilibrium. This assumption should be taken with
caution when studying transport close to the phase transition, as
typical time scales close to the phase transition diverge.

We perform the Gaussian integration over the field $\rho$ to obtain an effective action for $\Delta_T $.
This leads to an extra term in the $\Delta_T $ propagator, so that
\begin{eqnarray}
\nonumber
   -\frac{1}{\lambda}&+&\Pi_{\Delta_T\Delta_T}  \rightarrow \\
\nonumber
   && -\frac{1}{\lambda}+\Pi_{\Delta_T\Delta_T}+\frac{\left(\Pi_{\Delta_T\rho}\right)^2}{\left(2 V_c\right)^{-1}+\Pi_{\rho\rho}} \\
   &\approx&  -N_0\left(\frac{\pi}{4}\frac{D q^2}{\Delta_0}+\frac{\Omega^2}{4
   \Delta_0^2}\frac{1}{2AV_cN_0+1}\right).
\end{eqnarray}
Finally, the effective action describing real and imaginary parts of the
fluctuation in the order parameter is
\begin{eqnarray}\label{effective_action_re_im}
 \nonumber
  S_\mathrm{fluc} &=& N_0\int{\,\mathchar'26\mkern-9mu{\!d}}q\left\{\left(1+\frac{\Omega^2}{12 \Delta_0^2}+\frac{\pi}{8}\frac{Dq^2}{\Delta_0}\right)|\Delta_L(q)|^2\right.\\
   &&\!\!\!\!\!\!\!\!\!\left. \left(\frac{\pi D q^2}{4\Delta_0}+\frac{\Omega^2}{4
   \Delta_0^2}\frac{1}{2AV_cN_0+1}\right)|\Delta_T(q)|^2\!\!\right\}\!\!.
\end{eqnarray}
We note that the general expression for the \emph{uniform}
fluctuations of the magnitude of the order parameter, Eq.
(\ref{order_parameter_magnitude}), reduces to the form given by
$\Pi_{\Delta_L\Delta_L} $ in Eq. (\ref{effective_action_re_im}),
under the substitution $\Delta=\Delta_0+\Delta_L$. One can show
\cite{AvanOtterloCondMat1997} that the real and imaginary parts of
the order parameter, $\Delta_L/\Delta_0 $ and $\Delta_T/\Delta_0$,
are related under gauge transformation to the amplitude, $ \rho$
and phase, $\phi$ of the order parameter, respectively. Collecting
all terms and writing the action in terms of the amplitude and
phase, we have
\begin{eqnarray}
\nonumber
  S &=& N_0A\Delta_0^2\int_0^L dx\int_0^{1/T}
  d\tau\left\{\frac{\rho^2}{2}\left[\ln\left(\rho^2\right)-1\right]\right.\\
&&\left. +2\xi_0^2\rho^2\left[\phi'^2+\frac{\dot{\phi}^2
}{v_\phi^2}\right]+\xi_0^2\left[\rho'^2+\frac{\dot{\rho}^2}{v_{\rho}^2}
  \right]\right\},
\end{eqnarray}
with $ v_\phi$ and $v_\rho$ as given in Sec.~\ref{Low resistive
wires}.

\bibliographystyle {h-physrev3}

\begin{thebibliography}{10}
\bibitem{CommentBdGQuasiParticles}
This could be seen by considering a time dependent Bogoliubov de
Gennes (BdG)
  equation, with a time dependent phase: $\Delta =
  |\Delta|\exp\left(i\phi(t)\right)$. By the canonical transformation $(u',
  v')=\exp\left(\frac{i}{2}\sigma_z\phi\right)(u, v) $, we produce an effective
  potential $\dot{\phi}/2 $ for the quasi-particles in the BdG Hamiltonian.
\bibitem{CommentLAMHValidityWiskers}
This is in contrasted to the well established fit of the LAMH
theory to tin
  whisker crystals~\cite{RSNewbowerPRB1972}, where the theory is applicable for
  $1 \times 10^{-5}<\epsilon< 0.1 $, and the range of fitting was $1.1\times
  10^{-4}<\epsilon<3.5\times 10^{-4} $. Here $\epsilon = 1-T/T_c $ is the
  reduced temperature.
\bibitem{CommentDirtyLimit}
The expansion is done in the dirty limit, $\Delta_0\tau\ll 1$,
where $\tau$ is
  the scattering mean free time.
\end{thebibliography}
\newcommand{\noopsort}[1]{} \newcommand{\printfirst}[2]{#1}
\newcommand{\singleletter}[1]{#1} \newcommand{\switchargs}[2]{#2#1}
\providecommand{\bysame}{\leavevmode\hbox
to3em{\hrulefill}\thinspace}
\providecommand{\MR}{\relax\ifhmode\unskip\space\fi MR }
\providecommand{\MRhref}[2]{%
  \href{http://www.ams.org/mathscinet-getitem?mr=#1}{#2}
} \providecommand{\href}[2]{#2}

\end{document}